\documentstyle[11pt,newpasp,twoside,epsfig]{article}
\index{instructions}
\index{guidelines}

\def\cxo{{\em Chandra}}

\def\uu {4U\,0142$+$614}
\def\car {1E\,1048.1$-$5937}

\def\axj {AX\,J1844$-$0258}
\def\src {1RXS J170849$-$400910}
\def\ee {1E\,2259$+$586}

\newcommand{\BSAX}{{\em Beppo}SAX}

\def\edcomment#1{\iffalse\marginpar{\raggedright\sl#1\/}\else\relax\fi}
\reversemarginpar

\begin{document}
\title{Unveiling the multi-wavelength phenomenology of Anomalous
X--ray Pulsars} 

\author{GianLuca Israel and Luigi
Stella} \affil{INAF - Osservatorio Astronomico di Roma, Italy}
\author{Stefano Covino and Sergio Campana} \affil{INAF - Osservatorio
Astronomico di Brera, Italy} \author{Lorella Angelini}
\affil{NASA/Goddard, Space Flight Center, USA} \author{Roberto
Mignani} \affil{European Southern Observatory, Garching, Germany}
\author{Sandro Mereghetti} \affil{CNR, Istituto di Astrofisica
Spaziale e Fisica Cosmica, Milano, Italy} \author{Gianni Marconi}
\affil{European Southern Observatory, Paranal, Chile} \author{Rosalba
Perna} \affil{Harvard-Smithsonian Center for Astrophysics, Cambridge,
USA}

\begin{abstract}
During 2002-2003 the number of IR-identified counterparts to the
Anomalous X--ray Pulsars (AXPs) has grown to four (4U0142+614,
2E2259+584, 1E 1048-59 and 1RXS J170849-400910) out of the six
assessed objects of this class, plus two candidates.  More
importantly, some new common observational characteristics have been
identified, such as the IR variability, the IR flattening in the broad
band energy spectrum, the X-ray spectral variability as a function of
pulse phase (which are not predicted by the {\it magnetar} model), and
the SGR--like bursts (which can not be explained in terms of standard
accretion models).

We present the results obtained from an extensive multi-wavelength
observational campaign carried out collecting data from the NTT,
CFHT for the optical/IR bands, and XMM, Chandra (plus BeppoSAX
archival data) in the X-rays.  Based on these results and those
reported in the  literature, the IR-to-X-ray band emission of AXPs has been
compared and studied.
\end{abstract}

\section{Introduction}
It is now commonly believed that Soft $\gamma$--ray Repeaters (SGRs)
are magnetars --- neutron stars powered by their strong magnetic
fields (B$>$10$^{14}$\,Gauss). AXPs have been linked to SGRs because
of similar timing properties, namely large periods (P; in the 5-12s
range) and large period derivatives (\.P; Thompson \& Duncan 1993 and
1996). However, what differentiates these two seemingly dissimilar
objects is, at present, unclear. Nonetheless, there is a growing group
of radio pulsars (Camilo et al. 2000) with similarly long periods and
with inferred magnetic field strengths approaching $10^{14}$\,G.
These pulsars possess no special attributes linking them to either the
AXPs (no steady bright quiescent X--ray emission; Pivovaroff, Kaspi \&
Camilo 2000) or to the SGRs (no bursting history).  Thus periodicity
alone does not appear to be a sufficient attribute for
classification. Conversely, a very high magnetic field strength cannot
be the sole factor governing whether or not a neutron star is a
magnetar, a radio pulsar or in a binary system.
One possibility is that AXPs and SGRs are linked temporally.
Specifically, three out of the six AXPs are associated with supernova
remnants (SNRs) whereas only SGR\,0526--66 has a plausible SNR
association (Gaensler et al. 2001).  Taken at face value, these data
suggest that AXPs evolve into SGRs.  However, this hypothesis has at
least two severe problems (Kulkarni et al. 2003).  First, the
rotational periods of SGRs are similar to those of AXPs, about 10-s.
Second, inferred magnetic field strengths of SGRs are similar to (and
perhaps even larger than) those of AXPs (Mereghetti et al.  2002).

The recent detection of X--ray bursts from 1E\,2259+586 and
1E\,1048.1--5937 has strengthened the possible connection of AXPs with
SGRs (Kaspi \& Gavriil 2002; Gavriil et al. 2002). In the case of
1E\,2259+584, IR variability of the counterpart has been
detected few days after a strong X--ray bursting activity.  Also the
variability of the IR counterpart to 1E\,1048.1--5937 is thought to be
related to X--ray variability (Israel et al. 2002). Although these
new properties open a new horizon in the field, we do not understand
what specific physical parameter(s) differentiates AXPs from SGRs.

Evidence for flattening (or excess) of the flux in the IR band, with
respect to a simple blackbody component extrapolated from the X--ray
data, has been reported in four AXPs, namely 1E\,2259+586,
1E\,1048.1--5937, 4U\,0142+614 and 1RXS\,J1708--4009 (Hulleman et
al. 2001; Wang \& Chakrabarty 2002; Israel et al. 2003a and
2003b). The magnetar scenario (in its present form) does not account
for the observed IR emission or variability in AXPs and no predictions
can be, therefore, verified. On the other hand, the accretion models
fail in accounting for one of the main feature of SGRs, and recently
AXPs, that is the bursts.  In conclusion, none of the proposed
theoretical models (at least in their present form) seem to be able to
account simultaneously for the IR, optical and X--ray emission of
AXPs. In this respect the IR emission from AXPs/SGRs may play a key
role in the study and understanding of these sources.

\begin{figure}
\centerline{\psfig{file=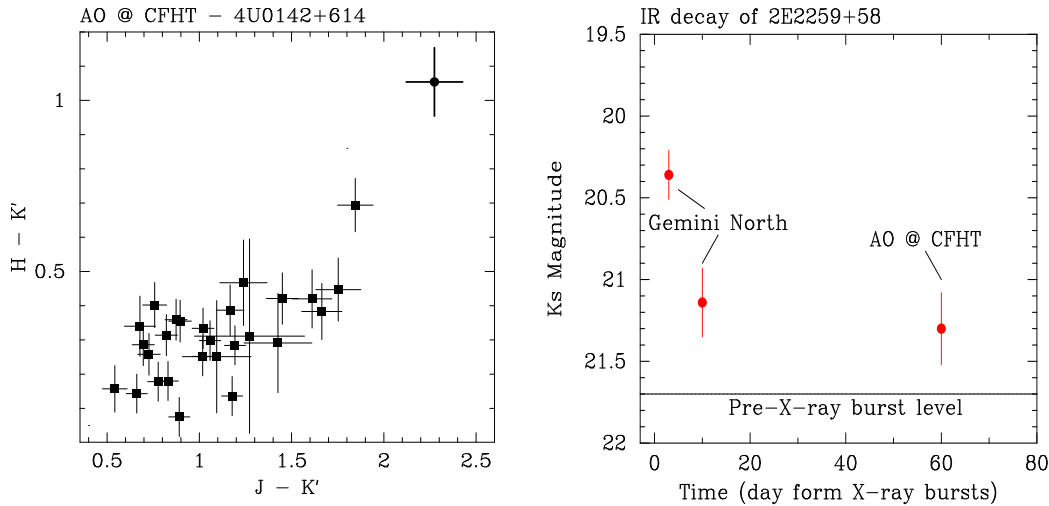,width=6cm}
\psfig{file=israelg_1b.ps,width=6cm}}
\caption{IR color--color diagram obtained for the region (radius of
30\arcsec) around the position of \uu\ and based on adaptive optics
observations carried out on August 2002 from the CFHT (left panel).
The counterpart (in the upper right corner) clearly stands out with
respect to the other objects in the field of view. IR decay ``lightcurve'' 
of \ee\ inferred by using the CFHT data presented here and those 
in literature (Kaspi et al. 2003; right panel). }
\end{figure}

\section{\cxo\ HRC--I and IR adaptive optics observations}
The IR-to-X--ray data we relied upon have been obtained as a part of a
joint ESO/\cxo\ large project aimed at the identification and study of
the optical/IR counterparts of AXPs in the southern sky, and the study
of the spatial distribution of X--ray emission from AXPs. About 10ks
\cxo\ observations have been obtained for \src\ and \car, while 20ks
have been devoted to \axj. In all cases the HRI--I instrument has been
used in order to obtain the most accurate X--ray position. 

Optical and IR data have been obtained for all the known AXPs from the
3.5-m New Technology Telescope (NTT, La Silla, Chile; SOFI and SUSI2
instruments), the 4-m Canada France Hawaii Telescope (CFHT with OAB),
and more recently from the Very Large Telescope (VLT with FORS and ISAAC).

The results obtained for \src\ and \car\ have been already presented
elsewhere and concern the identification of the likely IR counterpart
of \src, and the detection of IR variability from the IR counterpart
of \car\ (Israel et al. 2002, Israel et al. 2003a).

\uu\ has been observed at the CFHT in the J ($\sim$3000s effective
exposure time), H and K' ($\sim$4000s) bands, and thanks to the
adaptive optics we obtained a source PSF of $\sim$0\farcs16. After
standard photometric reduction we inferred the following magnitudes:
$J=22.3\pm0.1$, $H=21.1\pm0.1$ and $K'=20.0\pm0.1$ (Israel et
al. 2003b, in preparation). Figure\,1 (left panel) summarises the
photometric results; the IR counterpart of \uu\ clearly stands out in
the color-color diagram ($J-K'=2.3$) with respect to the other objects
in the field of view.
 \begin{figure}
\centerline{\psfig{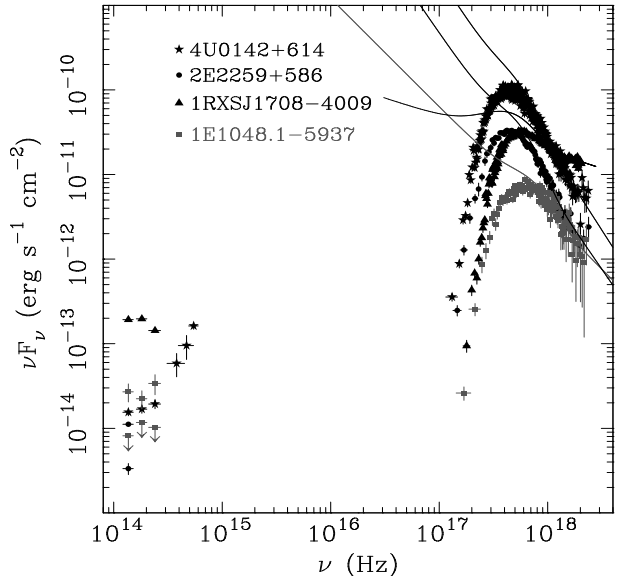}}
\caption{Broad band energy spectrum of AXPs.  X--ray raw data are
taken from the medium and low energy instruments on board \BSAX\ while
the solid upper curves are the unabsorbed fluxes for the black body
plus power law model. On the lower left corner of the plot are the
optical/IR unabsorbed fluxes.}
\end{figure}

K' band observations at the CFHT have been also carried out for 1E 2259+586
for a total exposure time of $\sim$7300s, 2 months after the detection
of SGRs-like burst activity from the X--ray pulsar (Kaspi \& Gavriil
2002). Also in this case, thanks to the relatively narrow PSF achieved
with the adaptive optics ($\sim$0\farcs19) we were able to detect the
IR counterpart at a K' magnitude level of 21.31$\pm$0.24 (about 0.4
magnitudes brighter than the magnitude reported by Hulleman et al 2001; 
see also Figure\,1, right panel). This result implies that after 2 months 
the IR ``activity'' of \ee\ is still present.  

Based on the above results and those already present in literature, we
have plotted the broad band energy spectra of the four AXPs for which
the IR/optical counterparts are known. In Figure\,2 we show the
IR-to-X--ray spectra (X--ray raw data are absorbed; solid lines
represent the unabsorbed models). 
It is worth noting the scatter in the X--ray unabsorbed fluxes that is
of a factor of about 10.  Also the IR fluxes show a relatively large
scatter of values. Moreover, we note that \src\ is by far the AXP
which most deviates from the other sources, both in the X--ray and in
the IR bands. Finally, the IR flux values of \uu\ presented here clearly
show a deviation from the black body component which was originally
used to fit the optical fluxes (Hulleman et al. 2000). Given the
extremely high values of the extinction in the direction of the AXPs,
we note that the flux ratio $F_X/F_{IR}$  have a large 
uncertainty, depending on the assumed X--ray energy interval.

\section{The debated case of \axj}
\axj\ is a candidate AXP since the discovery of 7s X--ray pulsations
in a 1993 ASCA dataset (Torii et al. 1998). The source was observed
once again with ASCA a year later and was detected at flux of a factor
of about 10 fainter. No periodicity was detected
(Vasisht et al 2000). Therefore, its association to the AXP class was
based merely on the period and spectral parameters of the source in
the ``high'' level.

In order to clarify the nature of \axj\ we requested 90ks-long
BeppoSAX, 20ks HRC--I \cxo, and 25ks XMM observations. Moreover we
obtained optical and IR images from the NTT. In all X--ray images we
detected only a relatively faint object, the position and count rates
of which were consistent with those of the second ASCA observation.
In Figure\,3 (left panel) we show the BeppoSAX and XMM spectra of
\axj\ which are well fitted with an highly absorbed ($N_H\sim
5\times10^{22}$ cm$^{-2}$) power law ($\Gamma=1.2\pm0.8$) or
alternatively a black body with a characteristic temperature of
$1.0\pm0.5$ keV. In both cases the unabsorbed luminosity is in the
$10^{33}-10^{34}$ erg s$^{-1}$ range (assuming 7 and 15 kpc as
fiducial distances). 

Due to the slightly off-axis position of the target in the \cxo\
pointing and to its faintness, the best position is a circular region
with a radius of about 1\arcsec. The optical/IR observations revealed
no peculiar objects within the uncertainty region down to a limiting H
magnitude of about 21. Spectroscopic VLT data of the closest object
to the \cxo\ position have been carried out and the analysis is in
progress end will be reported elsewhere.

\begin{figure}
\centerline{\psfig{file=israelg_3a.ps,width=6cm}
\psfig{file=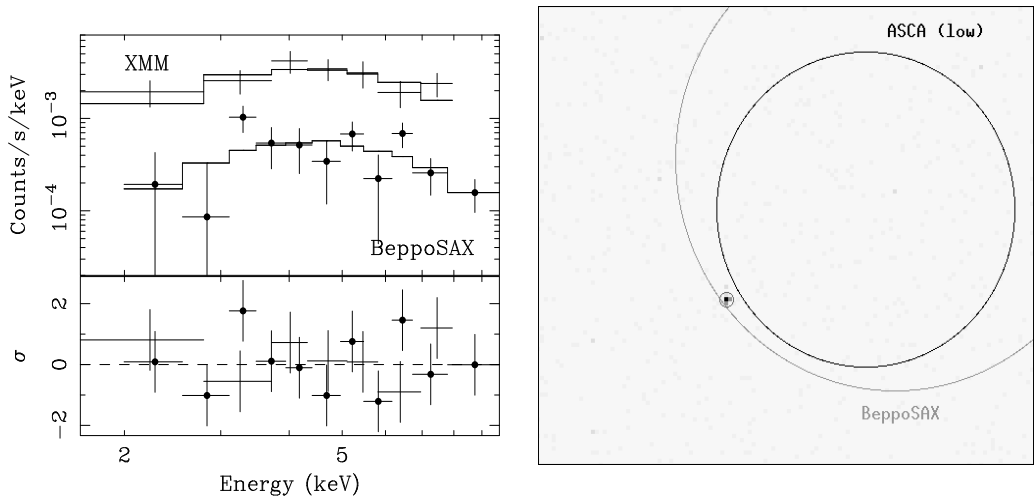,width=6cm}}
\caption{XMM EPIC and BeppoSAX MECS spectra of \axj\ fitted with an
absorbed power law model (left panel). The \cxo\ HRC-I image is shown
with superimposed the ASCA and BeppoSAX positional uncertainty regions
at the 90\% confidence level (20\arcsec\ and 35\arcsec, respectively;
right panel).}
\end{figure}

\acknowledgments This work is supported through CNAA, ASI, CNR and
Ministero dell'Universit\`a e Ricerca Scientifica e Tecnologica
(MURST) grants. The authors thank Olivier Hainaut, Leonardo
Vanzi and the NTT Team for their kind help during ESO observations. We
also thank Olivier Lai and the CFHT team for their support. We finally
thanks Hank Donnelly of the \cxo\ Team.

\end{document}